\documentclass[useAMS,usenatbib]{mn2e}

\usepackage{graphicx}
\usepackage{color}
\usepackage{multirow}
\usepackage{amssymb,amsmath}

\def\ps{$\rm km \,s^{-1}\,kpc^{-1}$}
\def\lv{\emph{l-v}}
\def\xy{\emph{x-y}}
\def\arcdeg{$^\circ$}
\def\kms{\,km\,s$^{-1}$}
\newcommand{\sech}{\mathrm{sech} \,}

\def\H2{$\rm H_2$}
\def\torus{\textsc{torus}}

\title[The morphology of the Milky Way - II.]{The morphology of the Milky Way - II. Reconstructing CO maps from disc galaxies with live stellar distributions}

\author[A. R. Pettitt, C. L. Dobbs, D. M. Acreman \& M. R. Bate]{Alex R. Pettitt$^{1,2}$\thanks{E-mail:
alex@astro1.sci.hokudai.ac.jp}, Clare L. Dobbs$^{2}$, David M. Acreman$^{2}$ \& Matthew R. Bate$^{2}$\\
$^{1}$Department of Physics, Faculty of Science, Hokkaido University, Sapporo 060-0810, Japan\\
$^{2}$School of Physics \& Astronomy, University of Exeter, Stocker Road, Exeter EX4 4QL, U.K.\\
}

\begin{document}

\date{\today}

\pagerange{\pageref{firstpage}--\pageref{lastpage}} \pubyear{2015}

\maketitle

\label{firstpage}

\begin{abstract}
{
The arm structure of the Milky Way remains somewhat of an unknown, with observational studies hindered by our location within the Galactic disc. In the work presented here we use smoothed particle hydrodynamics (SPH) and radiative transfer to create synthetic longitude-velocity observations. Our aim is to reverse-engineer a top down map of the Galaxy by comparing synthetic longitude-velocity maps to those observed. 
We set up a system of $N$-body particles to represent the disc and bulge, allowing for dynamic creation of spiral features. Interstellar gas, and the molecular content, is evolved alongside the stellar system. A 3D-radiative transfer code is then used to compare the models to observational data. The resulting models display arm features that are a good reproduction of many of the observed emission structures of the Milky Way. These arms however are dynamic and transient, allowing for a wide range of morphologies not possible with standard density wave theory. The best fitting models are a much better match than previous work using fixed potentials. They favour a 4-armed model with a pitch angle of approximately 20\arcdeg{}, though with a pattern speed that decreases with increasing Galactic radius. Inner bars are lacking however, which appear required to fully reproduce the central molecular zone.
}
\end{abstract}


\section{Introduction}
The question of the structure of the Milky Way is still an open one. A top-down view of our Galaxy has been sought after for decades, and yet due to our location within the disc itself such a map remains elusive.  The general consensus favours a multi-arm spiral pattern with some kind of inner bar structure, though the exact morphology and kinematics of these features are less well constrained \citep{2004ApJS..154..553S,2005AJ....130..569V,2008ASPC..387..375B,2012MNRAS.422.1283F,2014arXiv1406.4150P}. For example, spiral models with 2 or 4 arms have been proposed to best represent Galactic morphology \citep{1985IAUS..106..255E,1985IAUS..106..283L,1976A&A....49...57G,1993ApJ...411..674T,1970IAUS...38..126W} though there is evidence that the gaseous and stellar components trace out different arm numbers, with gas favouring a 4-armed structure and stars a 2-armed one \citep{2000A&A...358L..13D,2001ApJ...556..181D,2008ASPC..387..375B,2010ApJ...722.1460S}. 
Furthermore, the pitch angle of these arms also varies dramatically between studies, with a preference to tighter wound arms in the inner disc compared to the outer, with values ranging from 5\arcdeg{} to 20\arcdeg{} \citep{2006Sci...312.1773L,2009A&A...499..473H,2011ARep...55..108E,2012MNRAS.422.1283F}.

The status of the Galactic bar further complicates the already confusing picture. Values for the orientation, bar length, and rotation speed vary between studies, especially the rotation speed for which values range from 30\ps{} to 70\ps{} \citep{2011MSAIS..18..185G}. The suggestion of a second narrow bar in addition to the classical 3\,kpc long \emph{COBE} DIRBE bar \citep{1991ApJ...379..631B,1994ApJ...425L..81W,1997MNRAS.288..365B,2002ASPC..273...73G} only confuses matters further \citep{2000MNRAS.317L..45H,2005ApJ...630L.149B}.

\begin{figure}
\includegraphics[width=84mm]{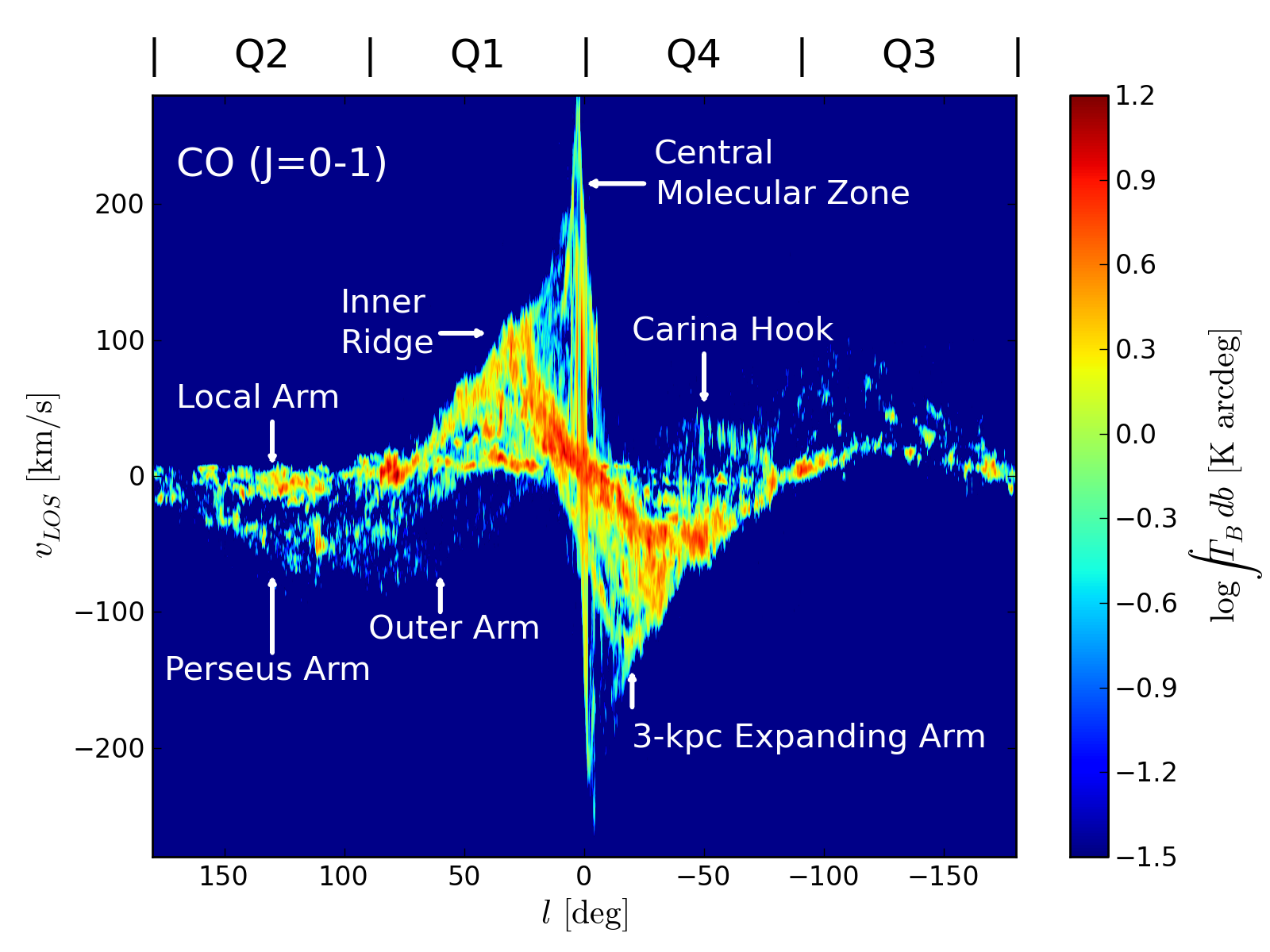}
 \caption{Longitude-velocity map of brightness temperature of the CO (J=0-1) transition (\citealt{2001ApJ...547..792D}), with major arm features labelled. We integrate the CO emission over latitude in order to show weaker features. Q1-4 indicates the position of Galactic quadrants.}
\label{DameLV}
\end{figure}

Determining the distances to sources to map out Galactic structure has many difficulties and uncertainties. 
Using longitude-velocity maps (\lv{}) can circumvent this issue by displaying spiral arm features without need of calculating distances. Such maps exist in a number of different sources \citep{2005A&A...440..775K,2007AJ....134.2252S,2012ApJS..199...12M,1987A&A...171..261C,1980ApJ...239L..53C,2001ApJ...547..792D,2006ApJS..163..145J,2013A&A...554A.103P}. The CO \lv{} emission map of \citet{2001ApJ...547..792D} is shown in Figure\,\ref{DameLV} where molecular gas can be seen to trace out the spiral structure as regions of increased emission strength.

Top-down maps of the Galaxy can be recreated by reverse-engineering models to match the features seen in maps such as Figure \ref{DameLV} (e.g. \citealt{1999MNRAS.304..512E}, \citealt{1999A&A...345..787F}, \citealt{2004ApJ...615..758G}, \citealt{refId0}, \citealt{2010PASJ...62.1413B}, \citealt{2013MNRAS.428.2311K}, \citealt{2015MNRAS.446.4186S}). In the first paper in this series, \citet{2014arXiv1406.4150P}, hereafter referred to as Paper 1, we used fixed analytical gravitational potentials for the arm and bar components in simulations of ISM gas to constrain Galactic structure. Synthetic longitude-velocity maps were constructed from simulations to compare directly with the observed CO emission. A wide array of potential morphologies were tested, including numerous pitch angles, pattern speeds and arm numbers. We found arm and bar features were necessary to match the observations, with arm pattern speeds of 20\ps{} and bar pattern speeds in the range of 50-60\ps{}. 
The 4-armed models could reproduce many of the features seen in observations but produced too much excess emission. The 2-armed model provided too few features compared to observations. The symmetric nature of such potentials, used to represent the spiral density waves of \citet{1964ApJ...140..646L}, made the exact placement of arms in \lv{} space problematic. For example, no 4-armed model could satisfactorily reproduce the Carina and Perseus arms simultaneously.

In this paper we take a different approach and model the stellar distribution of the Milky Way with a set of discrete $N$-body particles rather than a continuous potential as in Paper 1. This method has been used to simulate both isolated stellar systems \citep{1984ApJ...282...61S,1987MNRAS.225..653S,1988MNRAS.231P..25S,2010ApJ...720L..72S,2013A&A...553A..77G,2013ApJ...766...34D} and the simultaneous evolution of a gas disc \citep{1985ApJ...298..486C,1993A&A...272...37E,2006MNRAS.371..530C,2009ApJ...706..471B,2012MNRAS.426..167G,2011MNRAS.414.2498S,2013MNRAS.436.1836R,2013MNRAS.429.1949A,2013MNRAS.432.2878R}. 
Simulations such as these, and many others, have shown that both bar and spiral features are surprisingly easy to reproduce, though not ones that necessarily agree with theory. For instance, spiral arms do not appear to be steady spiral density waves such as those suggested by \citet{1964ApJ...140..646L}, and bars are seemingly overabundant, and difficult to reproduce with accompanying spiral structures. Instead, properties such as number of arms and pattern speeds are in better agreement with swing amplified instabilities (\citealt{1981seng.proc..111T,2013A&A...553A..77G}, see Section \ref{stabmode}).

This paper is organised as follows. In Section \ref{sec:numerics} we outline the numerical methods used in this paper and the setup procedure for the various Galactic components. In Section \ref{sec:results} we present the results of our simulations. This is divided into the hydrodynamical simulations (\ref{sec:reshydro} and \ref{othermodels}) and \lv{} maps constructed using a simple approximation (\ref{simpmapssec}) and a full radiative treatment (\ref{rtmapssec}). In Section \ref{discussion} we discuss the successes and limitations of this study, specifically with comparison to the analytical potentials of Paper 1, and we conclude in Section \ref{conclusions}.

\section{Numerical Simulations}
\label{sec:numerics}
We use the SPH code \textsc{sphNG} based on the original version of \citet{1990ApJ...348..647B}, but substantially modified as described in \citet{1995MNRAS.277..362B} and \citet{2007MNRAS.374.1347P}, and parallelised using both OpenMP and MPI. This is the same code as was previously used to study the spiral structure of M51 by \citet{2010MNRAS.403..625D}.

\subsection{Hydrodynamical simulations}
The stellar distribution is represented by a collection of $N$-body particles constituting the disc and bulge components, and in some cases the halo. 
Star particles have adaptive gravitational softening lengths, which are calculated in a similar manner to the smoothing lengths for SPH particles \citep{2007MNRAS.374.1347P}.
 
The gas disc itself is devoid of self-gravity, where we assume that the majority of the gravitational force stems from the stellar material. Gas self-gravity may dominate the smaller scales, but we are interested predominately in the global gaseous structure. 
As with Paper 1, stellar feedback and magnetic fields are not included. The lack of feedback will cause gas to collapse into the plane rapidly, though the inherent dispersion of the stars keeps the gas disc thicker than when using an analytic potential. 

A basic chemical network is also included, identical to that of Paper 1, in which the abundances of HI, H$_2$ and CO are evolved for each SPH gas particle. These abundances are then used as inputs to the construction of \lv{} maps. The gas is also subject to various heating and cooling effects \citep{2007ApJS..169..239G}.

\subsection{Initial conditions and Galactic potentials}
In our calculations we seed the initial positions and velocities of the star particles from a set of density profiles tailored to match the rotation curve of the Milky Way. The Galaxy is decomposed into halo, disc and bulge components, the density profiles for which are used to set the initial positions of the star particles by the tabulation and sampling of mass distribution functions. We initialise the velocities using the procedures from \citet{1993ApJS...86..389H}, also used by \citet{2009ApJ...706..471B}, which primarily requires integrating the separate moments of the collisonless Boltzmann equation. Velocities of the disc are circular with some dispersion, while the velocities of the halo and bulge are set in random orientations. 

Our various calculations are described in Table \ref{SPHNGruns}. 
Our four fiducial calculations are named Ba, Bb, Bc and Bd, and use a live stellar bulge and disc in a static halo potential to reduce computational expense (the a-d denoting the decrease in disc to halo mass ratio). The Db and Dc models do not contain a bulge component and as such are highly susceptible to bar formation. The Hb model is the same as Bb but with the replacement of a static halo with a live one.

\begin{table*}
\centering
 \begin{tabular}{@{}lc c| c| c c c c c c }
  \hline
  Calculation & $M_d [10^{10} \, \rm M_\odot]$ & $M_h [10^{10} \, \rm M_\odot]$ & $M_b [10^{10} \, \rm M_\odot]$  & $m(R_{\rm in})$ & $m(R_{\rm mid})$ & $m(R_{\rm out})$& Notes\\
  \hline
  \hline
  Ba & 5.3 & 44 & 1.05 & 2.0 & 2.8 & 4.0 &``Huge" disc mass \\
  Bb & 4.1 & 63 & 1.05 & 2.5 & 3.6 & 5.7 & ``Heavy" disc mass \\
  Bc & 3.2 & 83 &1.05 & 3.1 & 4.8 & 8.2 & ``Normal" disc mass \\
  Bd & 2.5 & 101 &1.05 & 4.0 & 6.5 & 11.5 & ``Light" disc mass \\
  \hline
  Db & 4.1 & 63 & - & 2.2 & 3.4 & 5.5 &  No bulge \\
  Dc & 3.2 & 83 & - & 2.8 & 4.6 & 8.0 &  No bulge \\
  \hline
  Hb & 4.1 & 63 &1.05& 2.3 & 3.3 & 6.4 &  Live halo \\
 \end{tabular}
 \caption{Description of different live-disc models. Refer to Figure \ref{fig:RCmass} for the resulting rotation curves for the main models. If the halo is not listed as live then it is represented by an analytic potential. The gas component of the disc is identical in all models with a total mass of $8\times 10^{9}\, \rm M_\odot$. The swing amplified spiral mode from Equation \ref{SwingAmp} is given at three radii. These are $R_{\rm in}=5{\rm kpc}$, $R_{\rm mid}=7.5{\rm kpc}$ and $R_{\rm out}=10{\rm kpc}$ corresponding to the inner, mid and outer disc.}
 \label{SPHNGruns}
\end{table*}

We choose a NFW profile \citep{1996ApJ...462..563N} to represent the dark matter distribution, which has a density profile of the form
\begin{equation}
\rho_{\rm h}(r)=\frac{\rho_{h,0}}{r/r_h(1+r/r_h)^2}
\end{equation}
where $r_h$ defines the halo viral radius and $\rho_{h,0}$ is the density scale factor and is given by
\begin{equation}
\rho_{h,0} = \frac{M_h}{4 \pi r_{200}^3} \frac{C_{NFW}^3}{\ln(1+C_{NFW})   +   C_{NFW}/(1+C_{NFW}) }.
\label{NFWrho}
\end{equation}
We use a fixed concentration and halo scale length for all calculations of $C_{NFW}=5$ and $r_{200}=122 \, \rm kpc$.
When using a live halo it is necessary to truncate the halo at some distance, where we truncate the halo distribution using a double exponential decay with a scale length of 50\,kpc.
The stellar disc is seeded from an exponential density profile \citep{1987gady.book.....B}
\begin{equation}
\rho_d(r,z) = \frac{M_d}{4 \pi R_d^2 z_d} \exp{(-r/r_d)}\sech^2{(z/z_d)}
\end{equation}
where $r_d$ and $z_d$ are radial and vertical scale length, which we set to $r_d=3.0\, \rm kpc$ and $z_d=0.3\, \rm kpc$.
The stellar bulge is represented by a spherical Plummer profile of the form
\begin{equation}
\rho_b (r)=  \frac{3M_b }{4\pi}   \frac{r_b^2}{  \left(r_b^2+r^2\right)^{5/2} }
\label{BulgeRho}
\end{equation}
which is also truncated similarly to the halo at a distance of 4kpc. The bulge scale length is $r_b=0.35\, \rm kpc$ and the mass is $M_b=1.05 \times 10^{10} \, \rm M_\odot$.

We set gas on similar orbits and positions to the star particles, only with much lower masses in accordance with a gas disc of the same mass used in Paper 1 for consistency ($8\times 10^{9}\, \rm M_\odot$). 
Gas was not distributed according to the observed surface density profile however, but rather in an exponential disc. The gas profile is driven by the stellar disc and settles after a few 10's of Myrs.

The resulting resolutions adopted for each set of computations is approximately 1 million disc star particles, and 0.1 million bulge particles.
A resolution of 3 million gas particles was used, a compromise between the 5 million used in Paper\,1 and 1 million. The latter was seen to be too weak to produce sufficient emission due to the poorer resolution of the high density and CO rich gas.

\begin{figure*}
 \includegraphics[trim = 0mm 0mm 0mm 0mm,width=150mm]{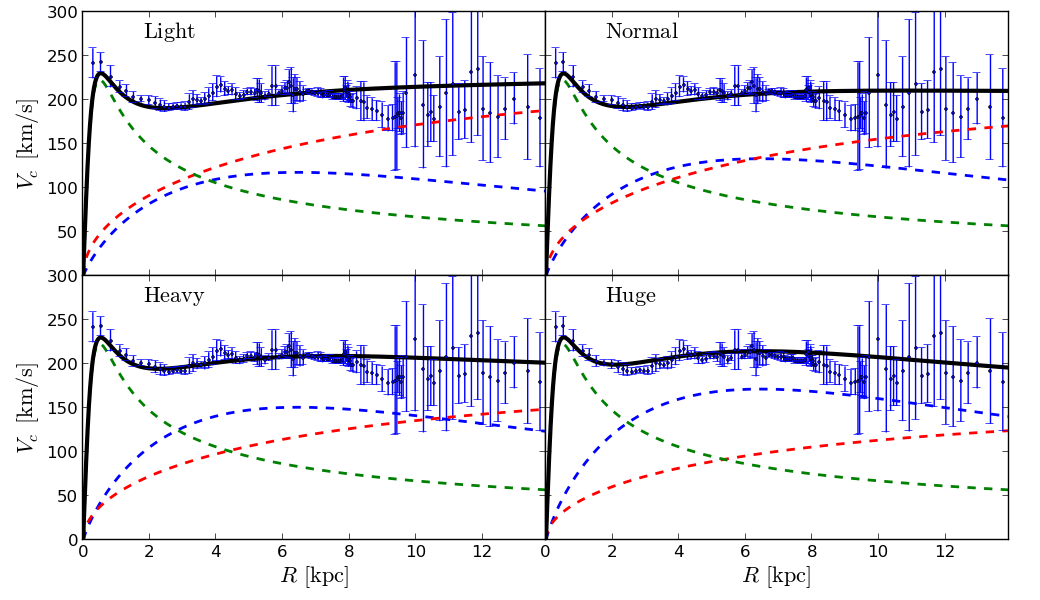}
  \caption{Axisymmetric rotation curves are shown for the calculations presented in this study. Disc, bulge and halo components are shown as the red, green and blue dashed lines respectively. Four different models are presented with labels referring to the disc mass; ``light", ``normal", ``heavy" and ``huge". The halo mass is adjusted in each case to ensure a rotation curve that is in agreement with observations (blue points, from \citealt{2012PASJ...64...75S}).}
  \label{fig:RCmass}
\end{figure*}

\subsection{Stability and spiral modes}
\label{stabmode}
The morphology of the stellar disc can be quantified using some simple parameters. The Toomre parameter in the stars, $Q_s$ \citep{1964ApJ...139.1217T}, characterises the stability of the disc to local collapse and is given by
\begin{equation}
Q_s = \frac{\kappa \sigma_R}{3.36 G \Sigma_0}
\end{equation}
where $\kappa$ is the epicycle frequency, $\sigma_R$ the radial velocity dispersion and $\Sigma_0$ the disc surface density. Values of $Q_s<1$ imply the disc is gravitationally unstable. Physically, $Q_s$ can be thought of as the balance between the disc pressure forces (driven by velocity dispersion in stars or thermal dispersion in gas) and gravitational attraction of the disc. The initial value of $Q_s$ must be defined for the disc to calculate the initial velocity dispersion. We use a value of $Q_s=1$ to ensure the disc is borderline stable to arm formation. 

The dominant swing amplified mode of the stellar disc, denoted $m$ \citep{1981seng.proc..111T}, can be calculated by
\begin{equation}
m=\frac{\kappa^2 R}{2\pi G \Sigma_0 X}\approx \frac{\kappa^2 R}{4\pi G \Sigma_0}
\label{SwingAmp}
\end{equation}
where $1<X< 2$ generates spiral features, and $X=2$ is a nominally adopted value \citep{2011ApJ...730..109F,2014PASA...31...35D}. Equation \ref{SwingAmp} shows that the number of arms is a strong function of the disc mass, with systems with high disc-to-halo mass ratios forming only a few strong spiral arms, whereas low mass discs form numerous but weaker arms \citep{1985ApJ...298..486C}. The dominant arm mode is also directly coupled to $Q_s$, and so seeding a disc with a very low $Q_s$ will lower the expected $m$. Equation \ref{SwingAmp} only predicts the dominant mode, and any single simulation will have other spiral modes of weaker or comparable presence. The value of $m$ also increases with radius, and so a simulation should show a greater number of arms in the outer disc compared to the inner disc.

As the disc mass drives the dominant spiral mode (Eq. \ref{SwingAmp}) we use four separate disc to halo mass ratios, the rotation curves for each are shown in Figure\,\ref{fig:RCmass} with the respective parameters for the ``light", ``normal", ``heavy" and ``huge" configurations (ordered by increasing disc mass) given in Table\,\ref{SPHNGruns}. We keep the bulge initial conditions the same for all calculations and use different disc and halo masses to reproduce the observed rotation curve. The values for the ``normal" setup are based on those from \citet{2009ApJ...706..471B}. 

The different disc masses also allow for different swing amplified spiral modes. While we do not have direct control over the spiral structures formed in these calculations, we attempt to drive a range of spiral modes induced by swing amplification. Table\,\ref{SPHNGruns} gives the predicted dominant spiral mode at three different radii corresponding to the inner, mid and outer disc (5, 7.5, and 10kpc). A wide range of arm numbers is predicted across all the models, with the Bd model aimed at being effectively flocculent.

\subsection{Emission maps}
We use the same method as outlined in Paper 1 to construct synthetic emission maps in CO using the radiative transfer code \torus{} \citep{Harries11072000}.
\torus{} creates cubes of brightness temperature as a function of latitude, longitude and line-of-sight velocity which are then integrated over latitude and quantitively compared to the CO map of \citet{2001ApJ...547..792D}. A turbulent velocity of 4\kms{} is added to the line width to better match the observed features. 

We must also take into account the uncertainty in the observer's position and velocity by varying the parameters $l_{\rm obs}$, $R_{\rm obs}$, $V_{\rm obs}$ (azimuthal and radial position and circular velocity of the observer in the disc). As the morphology is highly time-dependant we must also test multiple time-stamps for each simulation. We choose to do so in the range of 200-320Myrs of evolution of the gas disc. This is enough time to allow the majority of the molecular gas to form (found to be $t \gtrsim $150Myr) and for the disc to settle into a spiral pattern that will persist for up to a Gyr. Note that we must search the full range of $l_{\rm obs}$ values, as we cannot assume the disc is symmetric as in the grand design case of Paper 1. 

A simple method of creating \lv{} maps to narrow down the parameter space is used, which is described in detail in Paper 1. Maps in \lv{} space are made by projecting the gas from \xy{} to \lv{} space using $l_{\rm obs}$, $R_{\rm obs}$ and $V_{\rm obs}$, and an analogue for CO emission ($I_{i,\rm synth}$) is calculated using the distance from the observer to the gas ($d_i$) and the CO abundance ($\chi_{i,\rm CO}$) of the $i^{\rm th}$ SPH particle.
We have however slightly changed our equation for the synthetic emission map approximation by the addition of an $\epsilon$ term
\begin{equation}
I_{i,\rm synth} \propto \chi_{i,\rm CO}  / \left(d^k_i + \epsilon \right).
\label{synthLV}
\end{equation}
Because of the large amount of interarm material there are cases in our fitting where the $1/d_i^2$ approximation causes extremely large intensity due to particles very close to the observer location. This rarely happened in Paper 1, as molecular gas tended to be exclusively placed along the potential minima. To rectify this we include a limiting factor, $\epsilon$, in the denominator to limit emission from material near the Solar position. 
To find appropriate values of $\epsilon$ we constrain a simple map made from Eq. \ref{synthLV} to a map created from the \torus{} radiative transfer code, which gives a best fitting value of $\epsilon=1.0$. We maintain $k=2$ from Paper 1, finding it to be equally good a fit as previously.

At each main time-step after 190Myrs we created a number of simple \lv{} maps using Equation \ref{synthLV} and different values for the azimuthal position ($l_{\rm obs}$), radial position ($R_{\rm obs}$) and circular velocity ($V_{\rm obs}$) of the observer. A mean absolute error (MAE) statistic is used to quantify the difference between the maps and the data in Figure \ref{DameLV}. This is a sum over the pixels in a map of the absolute difference between brightness temperature of the synthetic and observed CO maps (see Paper 1 for a more detailed description of this process). The fit for several neighbouring time-steps is compared, providing a best-fitting time, $t_{bf}$, for each model. The radiative transfer maps built by \torus{} are then used to quantitively compare different models and the CO observations.

\section{Results}
\label{sec:results}
We devote the majority of this section to discussion of our fiducial set of models. The SPH simulations and top-down maps are presented first, and both kinds of \lv{} maps after.

\subsection{Simulations and morphology of fiducial models}
\label{sec:reshydro}

\begin{table*}
\centering 
 \begin{tabular}{@{}l | c c c c c c c}
  \hline
  Model & $N(R_{\rm in})$ & $N(R_{\rm mid})$ & $N(R_{\rm out})$ & $\Omega_p(R_{\rm in})$ [\ps] & $\Omega_p(R_{\rm mid})$ [\ps] & $\Omega_p(R_{\rm out})$ [\ps] & $\alpha$ [\arcdeg]\\
  \hline
  \hline
  Ba & \textbf{2} & 5 & 3 & 38 & 44 & 33 & 19\\
  Bb & 2 & 4 & \textbf{4} & 34 & 25 & 21 & 20\\
  Bc & 4 & \textbf{4}  & 4 & 37 & 27 & 25 & 23\\
  Bd & 2 & 4 & \textbf{5} & 39 & 35 & 28 & 22\\
  \hline
  Db   & \textbf{3} & 3 & 5  & N/A & N/A & N/A & 27\\
  Dc   & 3 & \textbf{5} & 5 & N/A & N/A & N/A & 28\\
  Hb & 3 &  \textbf{4} & 4 &42 & 29 & 33 & 26\\
 \end{tabular}
 \caption[]{
The number of arms, pattern speed and pitch angle for each model. The number of arms and pattern speed are shown are three different radii as they are both a function of radius. The three radii of calculation are $R_{\rm in}=5{\rm kpc}$, $R_{\rm mid}=7.5{\rm kpc}$ and $R_{\rm out}=10{\rm kpc}$, corresponding to the inner, mid and outer disc. Boldfaced arm modes denote the dominant mode and the arm number corresponding to the $\alpha$ fit.}
 \label{SPHNGmorphFits}
\end{table*}

\begin{figure*}
  \centering
  \resizebox{1.0\hsize}{!}{\includegraphics[trim=0cm 0.0cm 5cm -1cm]{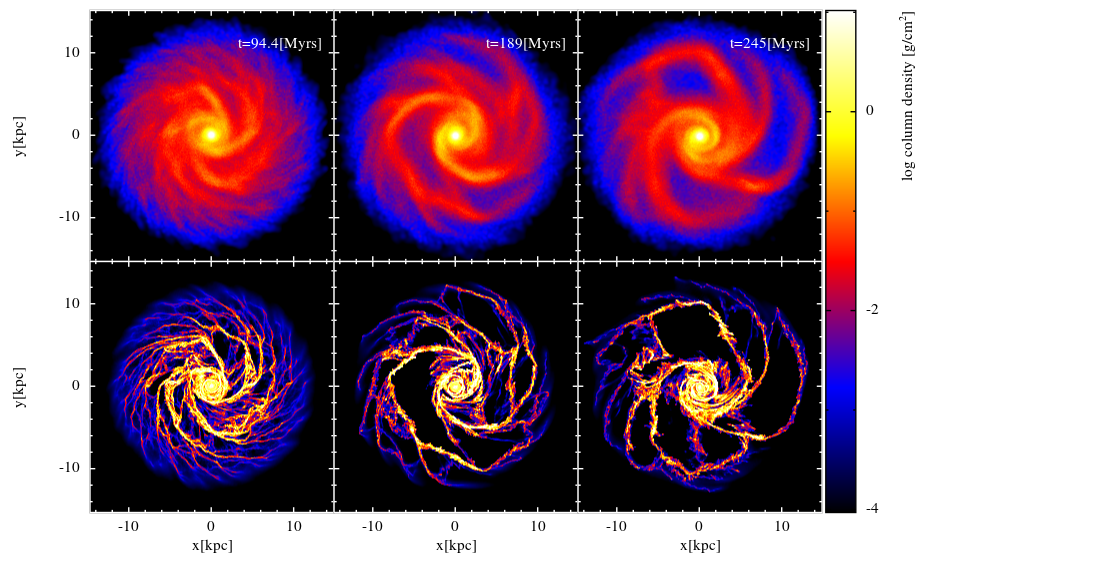}}
    \caption[Evolution of ``huge" disc model]{Evolution of the Ba model, where the heaviest disc mass is investigated, showing the stellar (top) and gaseous component (bottom). Large scale arm structures can be seen, yet appear rather irregular with many knee and kink features.}
    \label{fig:BulgeHuge}
  \centering
  \resizebox{1.0\hsize}{!}{\includegraphics[trim=0cm 0.0cm 5cm -1cm]{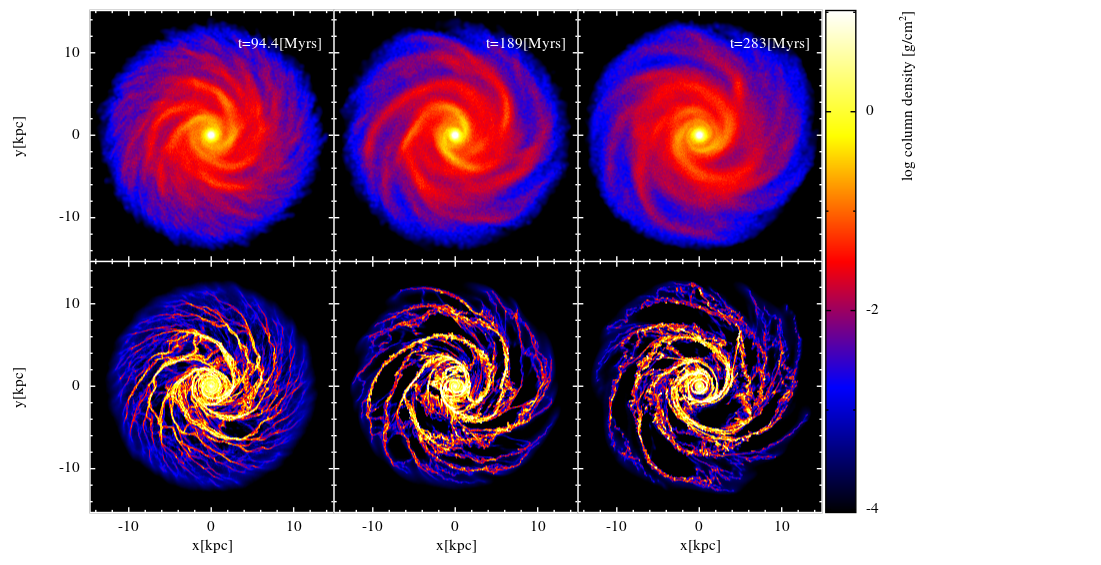}}
    \caption[Evolution of ``heavy" disc model]{As Fig. \ref{fig:BulgeHuge} but for the Bb configuration. Arms appear more smooth and regular compared to the Ba model.}
    \label{fig:BulgeHeavy}
\end{figure*}

\begin{figure*}
  \centering
  \resizebox{1.0\hsize}{!}{\includegraphics[trim=0cm 0.0cm 5cm -1cm]{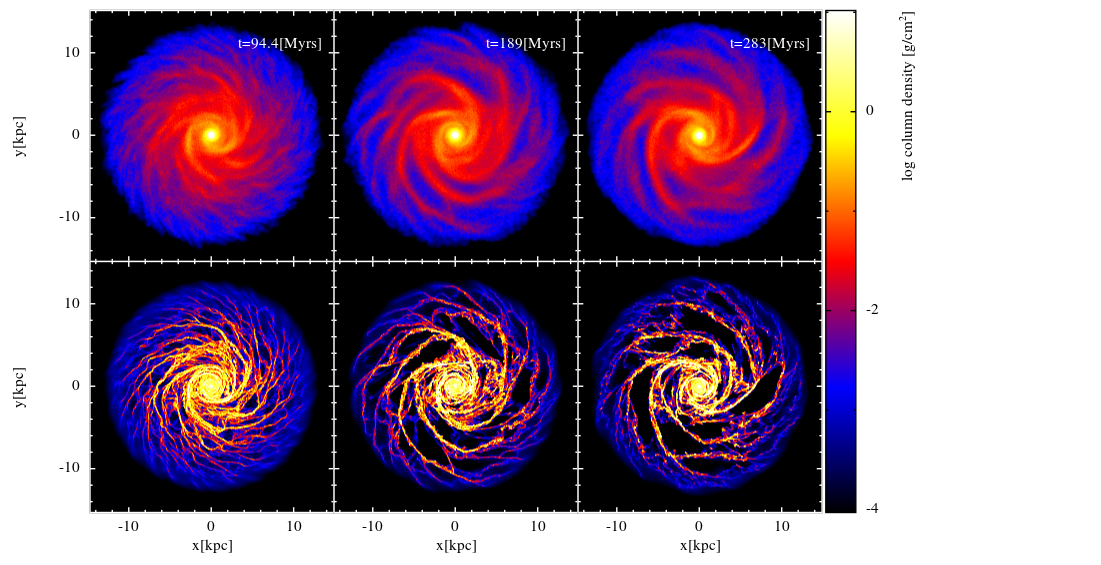}}
    \caption[Evolution of ``normal" disc model]{As Fig. \ref{fig:BulgeHuge} but for the Bc configuration. A significant number of small arm structures are now visible compared to Ba and Bb.}
    \label{fig:BulgeNorm}
  \centering
  \resizebox{1.0\hsize}{!}{\includegraphics[trim=0cm 0.0cm 5cm -1cm]{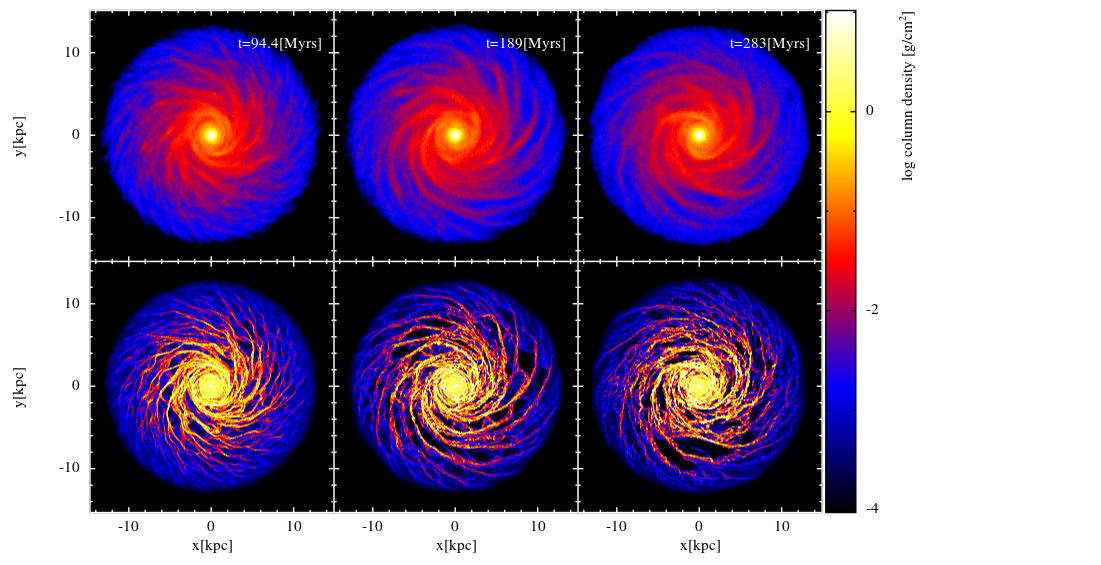}}
    \caption[Evolution of ``light" disc model]{As Fig. \ref{fig:BulgeHuge} but for the Bd configuration. The arm structure is very small scale and flocculent with no clear dominant spiral mode visible by-eye.}
    \label{fig:BulgeLight}
\end{figure*}

The main Galactic disc configuration investigated is the Ba-Bd group of models, which contain a live disc and bulge, but maintain a static dark matter halo.
We show the Ba, Bb, Bc and Bd models at three separate intervals in Figures \ref{fig:BulgeHuge}, \ref{fig:BulgeHeavy}, \ref{fig:BulgeNorm} and \ref{fig:BulgeLight} respectively. Stars are shown in upper panels, and gas in the lower. The third time-frame of the Ba model is slightly earlier than the other models due to the computation taking significantly longer than the others\footnote{Run time appeared to scale with stellar disc mass due to time-step limiting based on the accelerations of the gas particles moving from very low to high density.}. 
We did not continue this calculation further as maps made from existing time-steps and lower resolution calculations that had evolved for longer showed the Ba model was a poorer fit to Galactic \lv{} data in general.
For each model we attempt to quantify the arm number ($N$), pitch angle ($\alpha$) and pattern speed ($\Omega_p$), the methods for which are described in the Appendix. 

The gas clearly traces the arm features in the stars in Figures \ref{fig:BulgeHuge}, \ref{fig:BulgeHeavy}, \ref{fig:BulgeNorm} and \ref{fig:BulgeLight}, with the higher density gas in the disc being the location of the majority of the molecular content. In the highest mass disc (Ba) there is very little inter-arm structure, with large regions devoid of gas (Fig. \ref{fig:BulgeHuge}). In general the arm/inter-arm contrast is much stronger in the gas than in the stars. 
The arms in the Ba model appear non-logarithmic in many places, especially so in the final panel, with various kinks, or knees, forming along the spiral arms. These irregular structures also exist in other models, but are strongest in Ba. The spiral structure is much more regular in the moderate mass discs; Bb and Bc. 

The individual Fourier power spectra, fit to pitch angle and pattern speeds for each model are also given in the Appendix, with the best-fitting values presented in Table \ref{SPHNGmorphFits}. We give $N$ and $\Omega_p$ at three different radii, an inner, mid and outer region of the disc. The pitch angle is only quoted at a single radius as many $R$-$\theta$ pairs are required for the determination. The values in Table \ref{SPHNGmorphFits} are calculated at the time where the \lv{} map best-matches the Dame CO map (see Sec. \ref{simpmapssec}).
The arm numbers show a clear correlation between dominant arm mode (in bold) and disc mass. The agreement between the dominant value of $N(R)$ and the inner-mid $m(R)$ values (Table \ref{SPHNGruns}) is good, but the direct comparison at specific radii is poorer. We infer that Equation \ref{SwingAmp} is a good rough estimate of the expected dominant arm number in a simulation, but cannot be expected to exactly reproduce the dominant mode at individual radii.

In all of the live disc-bulge models the dominant arm number is seen to increase with radius, similar to the radial dependance predicted by Equation \ref{SwingAmp}. In Bb for example there is a strong inner 2-armed structure, and a strong 4-armed component in the outer disc (see the Fourier spectrum of Fig. \ref{ArmModes} and the top down maps in Fig. \ref{fig:BulgeHeavy}). The Milky Way is thought by some to have a higher arm number in the outer disc, and there is also confusion over whether there are 2 or 4 spiral arms \citep{1997MNRAS.286..885A,2006Sci...312.1773L}. The Bc model has a weaker radial dependance, with the $m=2$ mode being weaker in the inner disc compared to Bb. It does however have considerable power in the $m=5$ mode at around the Solar position. This is interesting as in many Galactic models the preferred structure is a 4-armed spiral, but with the addition of some small spur or minor Local arm which is always separate from the 4 main arms. The fact that the Bc model is preferentially a 4-armed model throughout most of the inner disc, lightly 2-armed in the centre, and seemingly 5-armed near the Solar radius makes it an excellent candidate for replicating Galactic \lv{} features. 

The live disc-bulge model with the lightest disc (Bd) appears to be a clear departure from the Bb and Bc models in that there is no clear dominant spiral mode visible by-eye. The structure of this disc appears to mimic a flocculent spiral galaxy such as NGC\,4414 with multiple small scale arms and inter-arm features \citep{1993A&A...280..451B}. What structure there is favours a $m=2$ mode in the inner disc, similar to the other models, and a $m=5$ mode in the outer disc.

The pitch angles have a weak correlation to disc mass, with an approximate value across all models of 21\arcdeg{}. This value is within agreement of the values suggested by the literature the Milky Way (e.g. \citealt{2006Sci...312.1773L}), though somewhat on the larger end of most estimates. The pitch angle for the lightest disc was not well constrained due to the highly flocculent nature of the disc. In all cases the logarithmic structure is best fit in the inner/mid-disc. 

The pattern speeds in Table \ref{SPHNGmorphFits} clearly decrease with radius for all models (see also Fig. \ref{ArmSpeeds}). The arms generated here are therefore material in nature, and are continuously sheared out and re-formed during rotation. The dominant arm number tends to be maintained, with arms being rebuilt from the remnants of the wound-up arm. The resulting pattern speeds range from around 40-35\ps{} in the inner disc to 30-20\ps{} in the outer disc, generally greater than those values suggested for the arms in previous work, centring around $\Omega_p\approx 20$\ps{} (e.g. \citealt{2005ApJ...629..825D,2011MSAIS..18..185G,2014arXiv1406.4150P}).

\begin{figure}
\includegraphics[trim = 0mm 3mm 0mm 10mm,width=84mm]{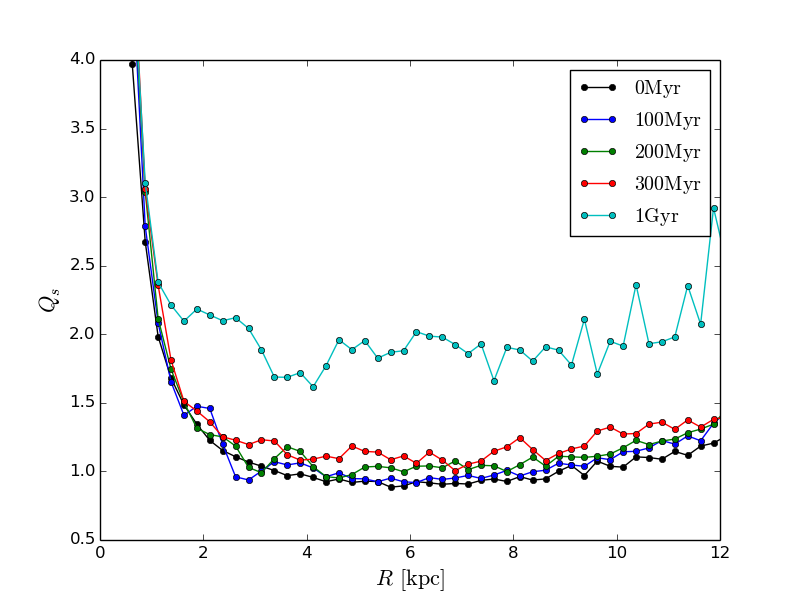}
 \caption{Evolution of the Toomre $Q_s$-parameter in the Bc disc-bulge calculation as a function of radius initially and after 0.1, 0.2, 0.3 and 1Gyr of evolution.}
\label{fig:BulgeQ}
\end{figure}

In Figure \ref{fig:BulgeQ} the evolution of $Q_s$ is shown from 0-1Gyr for the Bc model. Initially $Q_s\approx 1$ in the mid-disc, as dictated by the setup conditions, making the disc borderline stable. Over 300Myrs $Q_s$ can be seen to slowly increase throughout the disc, raising to approximately 1.2 in the mid-disc region. At much later times, of the order of Gyr, $Q_s\approx 2$ implying the disc is highly stabilised. The spiral structure of the various discs tends to prevail up to a Gyr, but begins to smooth out and dissipate at later times. 
The longevity of the spiral arms is also seen to be directly tied to the resolution of the stellar disc \citep{2011ApJ...730..109F}, with simulations with only $1\times 10^5$ particles displaying clear spiral arms for less than a Gyr. 
The issue of spiral longevity is a standing problem, and not one we aim to investigate here (see instead \citealt{1984ApJ...282...61S,2011MNRAS.410.1637S,2011ApJ...730..109F,2013ApJ...763...46B,2013ApJ...766...34D}).

The rotation curves of each live disc-bulge model show a dispersion that increases with disc mass. 
Dispersion around the mean rotation curve ranges from $\pm 50$\kms{} in Ba to $\pm 20$\kms{} in Bd. The dispersion in the rotation curve therefore appears directly related to the number of arms formed, which in unsurprising as the dispersion is also incorporated into the value of $m$ predicted by swing amplification ($m\propto\Sigma^{-1}$, and $\Sigma$ determines the stellar velocity dispersion). 

\subsection{Simulations and morphology of other models}
\label{othermodels}
\begin{figure*}
  \centering
  \resizebox{1.0\hsize}{!}{\includegraphics[trim=0cm 0.0cm 5cm -1cm]{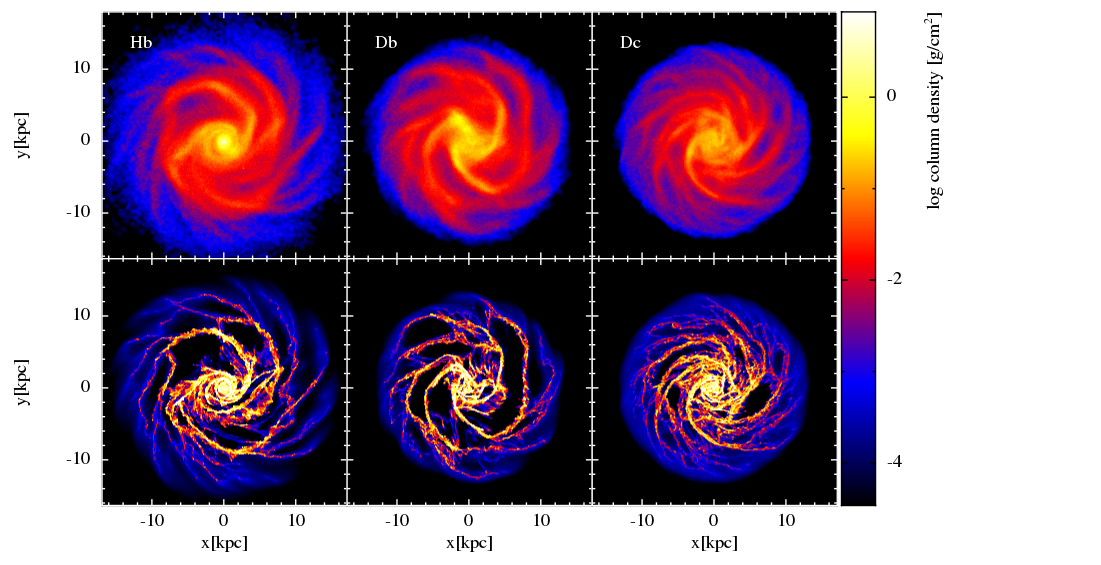}}
    \caption{Structure of three separate Milky Way models in stellar (top) and gaseous (bottom) components after 250Myrs of evolution. The models are a live a disc-bulge-halo system (Hb, left) and a live stellar disc without an inner bulge with a moderate (Db, middle) and light mass disc (Dc, left).}
    \label{fig:OtherModels}
\end{figure*}

In addition to our fiducial live bulge and disc models we performed additional calculations to test how omitting a bulge, and adopting a live instead of fixed halo modify our results. This included two models with no bulge component and two different disc to halo mass ratios (Db and Dc), and a model with a live halo (Hb). In Figure \ref{fig:OtherModels} we show the Hb, Db and Dc models in stellar (top) and gaseous (bottom) components after 250Myrs of evolution.

The structure of the Db and Dc models is similar to the Bb and Bc in the outer disc, but irregular in the centre without the inclusion of a bulge, showing a strong $m=3$ mode in the inner disc. The dominant arm modes for Db and Dc are $m=3$ and $m=5$ respectively, though both are present in each model (Table \ref{SPHNGmorphFits}). The pitch angle however gives very high values, with values of around 27\arcdeg{} for the Db and Dc models, far outside values inferred for the Milky Way and at the high end of values seen in external galaxies. The pattern speed is difficult to determine, as the arm structures in the mid/inner disc wind up very quickly due to lack of support provided by the inner bulge.

Due to the nature of the bulge-free system, the rotation curve for these models decays rapidly approaching the Galactic centre. This leads to the lack of an inner Lindblad resonance, which is believed to encourage the growth of $m=2$ modes (i.e. a bar) due to the lack of a $Q$-barrier in the inner disc \citep{1987gady.book.....B,1995gaco.book.....C,2014PASA...31...35D}. If we allowed the bulge-free models to evolve for the order of several Gyr then an inner bar component is seen to develop, but at the expense of arm structure.

For the live halo-disc-bulge model, Hb, the morphology is not dissimilar to the static halo models, specifically appearing somewhere between Ba and Bb in terms of arm features. 
The difference in morphology between this and the Bb model is likely caused by the additional dynamical flexibility caused by the resolved halo, which may be under-resolved to perfectly match the analytical potential, as well as the added errors induced by the artificial radial truncation. There appears to be a dominant 3-armed pattern in the inner/mid disc, which dissipates in the outer disc where a $m=4$ mode then dominates. The pattern speed for the $m=4$ mode gives pattern speeds in the range 30\ps{}$\le \Omega_p \le 45$\ps{}. 

\subsection{Simple kinematic \lv{} maps}
\label{simpmapssec}

\begin{table}
\centering
 \begin{tabular}{@{}l | c c c c}
  \hline
  Model & $t_{\rm bf}$ [Myr] & $R_{\rm obs}$ [kpc] & $V_{\rm obs}$ [\kms] & Fit stat.\\
  \hline
  \hline
  Ba & 197 & 8.5 & 215 & 0.994\\
  Bb  & 226 & 8.5 & 200 & 0.857\\
  Bc & 207 & 7.0 & 200 & 0.833\\
  Bd & 207 & 7.0 & 205 & 0.768\\
  \hline
  Db   & 235 & 8.5 & 200 & 0.974\\
  Dc   & 216 & 8.0 & 200 & 0.931\\
  Hb & 216 & 8.0 & 205 & 0.925\\
 \end{tabular}
 \caption{Results of fitting to $l_{\rm obs}$, $V_{\rm obs}$, $R_{\rm obs}$ and $t_{\rm bf}$ for each model. A lower fit statistic signifies a better fit to the observational data of Figure \ref{DameLV}.}
 \label{SPHNGsimpFits}
\end{table}

\begin{figure*}
 \includegraphics[trim = 0mm 0mm 20mm 0mm,width=160mm]{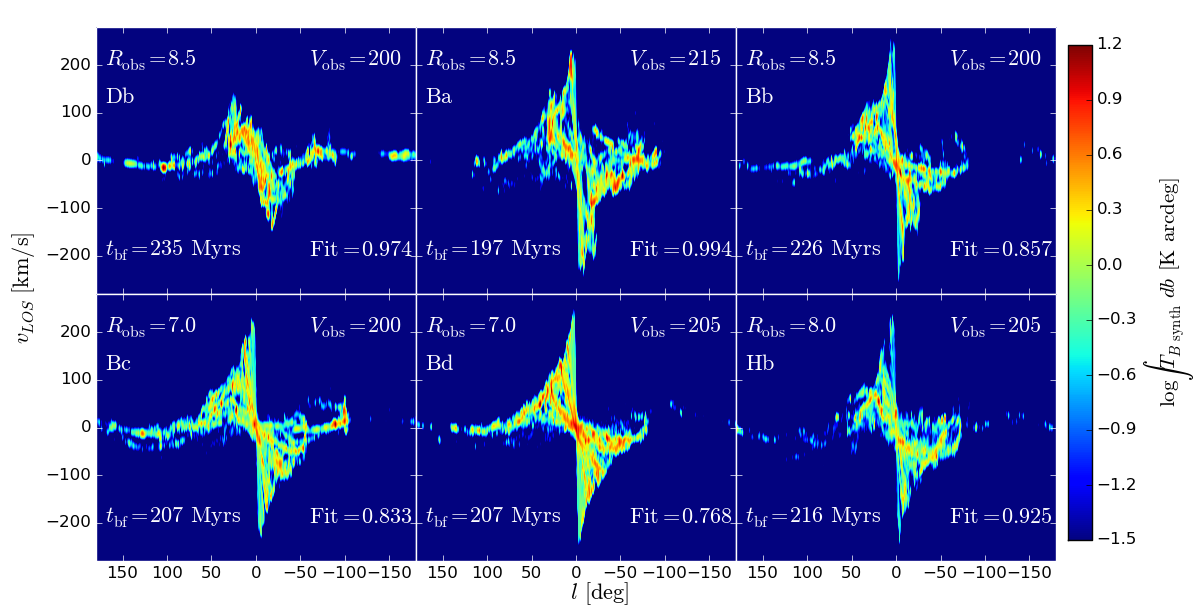}
  \caption{Best fitting \lv{} maps are shown from the simple fitting procedure to find $l_{\rm obs}$, $V_{\rm obs}$, $R_{\rm obs}$ and $t_{\rm bf}$ for models Db, Ba, Bb, Bc, Bd and Hb. Best fitting values are given in each panel, along with the value for the fit statistic used to constrain them. $V_{\rm obs}$ and $R_{\rm obs}$ are in \kms{} and kpc respectively. The  brightness of the emission has been scaled to the range seen in observations, whereas that in Figure\,\ref{MapLabel} has been independently calculated.}
  \label{fig:LVsimpleAll}
\end{figure*}

The process of fitting to $l_{\rm obs}$, $V_{\rm obs}$, $R_{\rm obs}$ and $t_{\rm bf}$ against the CO observational data of Figure \ref{DameLV} was performed for each model, where a lower fit statistic signifies a better fit to the observational data of Figure \ref{DameLV}. The resulting best fit \lv{} maps are shown in Figure \ref{fig:LVsimpleAll} with parameters for each model given in Table \ref{SPHNGsimpFits} (Dc is omitted from the figure as it differs little compared to Db). The bulge-free models (Db, Dc) have some of the poorest fit values, which appears to be due their lack of emission at high velocities in the inner disc. Some arm structures are seen in the outer disc, but these do not stray far from the local velocity ($v_{los}\approx 0$\kms{}). 

The remaining models, all with a live bulge component, provide a variety of \lv{} features. The heaviest disc model (Ba) appears to have a large velocity dispersion in the inner disc. While it has an inner structure that is aligned similarly to observations, there is a large amount of material at high velocities and not matching the observed features. Coupled with the irregular arm structure seen in the top-down map, we conclude this (our heaviest disc) is a poor match to the Milky Way and do not produce any full radiative transfer \lv{} maps. Moderate-to-light mass discs (Bb, Bc, Bd and Hb) provide a better agreement (and lower fit statistics). The lightest disc produces a near-uniform emission structure in the inner disc due to the flocculent nature of the arms. The Bb and Bc models provide good representations of the Carina and Local arms, while producing an Inner Ridge of the correct orientation. 

\subsection{Best fitting models and radiative transfer \lv{} maps}
\label{rtmapssec}

We created synthetic emission maps using the radiative transfer code \torus{} for the models at the best-fitting time-frame found in Section \ref{simpmapssec}. 
The fit statistic has been re-calculated from the \torus{} models now that the intensity correctly takes into account emission and absorption effects. The fit statistic follows the same trends across the maps as when we used the simple map creation tool. The three best fitting maps (in order) are the Bd, Bc and Bb models. In the case of the Bc model there was a later time-frame that had almost as good a fit (0.846 on the scale of the fit statistics in Figure \ref{fig:LVsimpleAll}), almost 100Myrs later at 292Myrs, the maps for which we also include here.

\begin{figure*}
 \includegraphics[trim = 0mm 0mm 20mm 0mm,width=160mm]{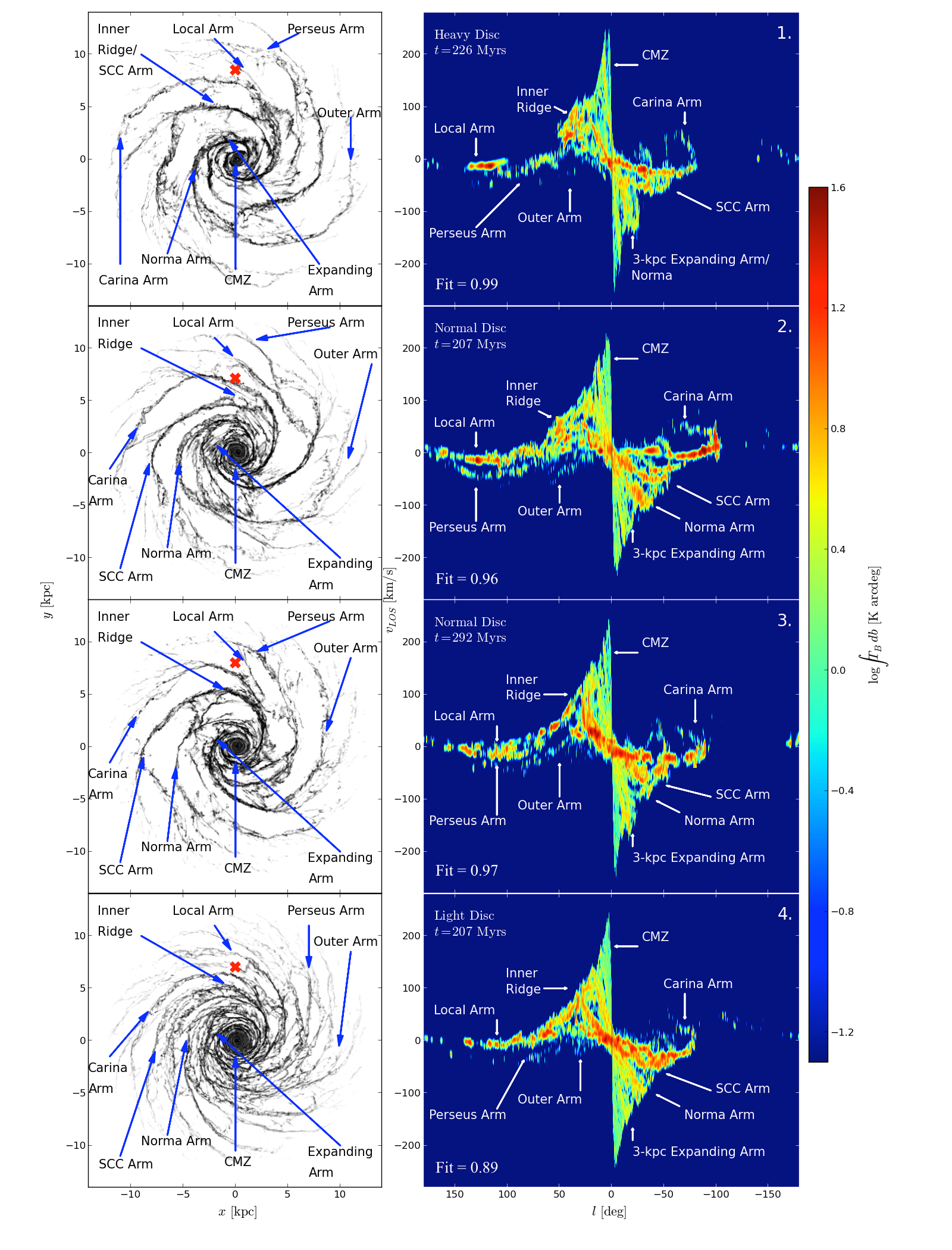}
  \caption{Our four best-fitting CO radiative transfer \lv{} emission maps with their \xy{} counterparts. The models from top to bottom are Bb, Bc(207 Myrs), Bc(292 Myrs) and Bd. The top-down maps only show gas material that is seen in CO \lv{} space; that of the highest density. The cross indicates the observer's position (which differs between models). SCC refers to the Scutum-Centaurus-Crux arm in the 4-armed paradigm of the Milky Way, also referred to in the main text as the Inner Ridge when viewed in \lv{} space. Arrows indicate locations of prominent features in \lv{} space.}
  \label{MapLabel}
\end{figure*}

In Figure\,\ref{MapLabel} we show our four best maps, from top to bottom: Bb, Bc(207Myrs), Bc(292Myrs) and Bd. In the left column we show the top-down gas distribution, and in the right the \torus{} CO \lv{} emission maps, created at the best fitting values of $l_{\rm obs}$, $V_{\rm obs}$ and $R_{\rm obs}$. In each frame we also label significant arm features, with the same nomenclature as used for the Milky Way (analogous to Fig. 25 in Paper\,1).

In the heavy disc (Bb, top panel) model the inner \lv{} features are smiler to the observed Inner Ridge, though not as clearly defined. The top-down map shows this is a combination of a far and near arm feature. The near arm (labelled Carina) appears weaker than the far inner arm (labelled SCC; Scutum-Centaurus-Crux). This allows for the reproduction of the Carina Arm in \lv{} space, which requires an arm to be very close to the observers location, while avoiding large amounts of emission at local velocities in the range $|l|<50^\circ$ as Carina Arm passes the line-of-line to the Galactic Centre. This model also has a feature at very low velocities similar to the 3kpc-Expanding Arm, though is angled steeper in \lv{} space than observations. A Local Arm feature has been produced by a spur of the Carina Arm, lying very close to the red-cross in the top-down map. This has been suggested by other studies; that the Local Arm is in fact some spur or inter-arm structure, rather than a primary arm (though the picture is still not clear; \citealt{2009ApJ...700..137R}, \citealt{2013ApJ...769...15X}). The caveat of this model is the outer arm structure. Both the Perseus and Outer arms, while clear in the top-down map, are weak or incorrectly placed in \lv{} space. The Perseus arm appears at velocities too similar to the local values, making it nearly indistinguishable from the Local Arm in CO emission. The Outer Arm can barely be seen in emission, its presence only given away by a couple of dense pockets of gas.

\begin{figure*}
\centering
\resizebox{0.8\hsize}{!}{\includegraphics[trim = 0mm 0mm 0mm 0mm]{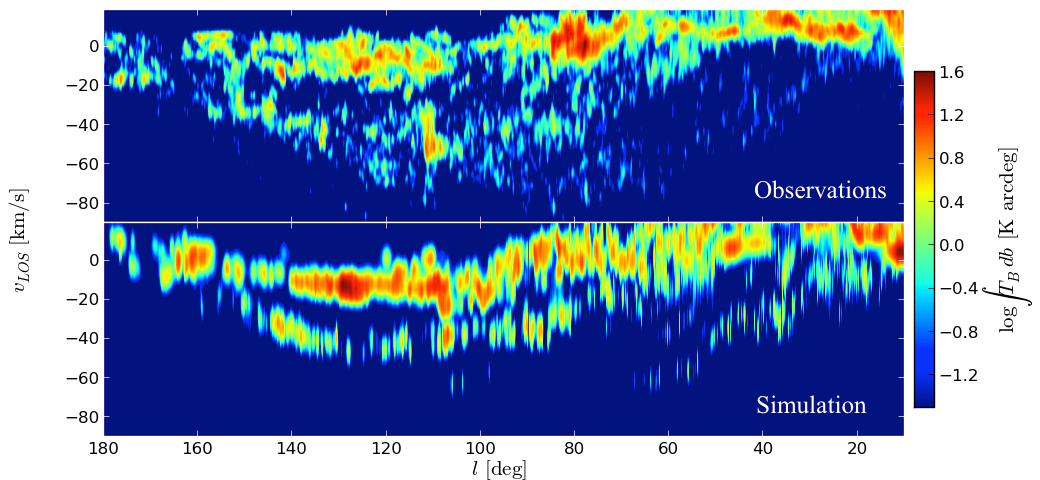}}
  \caption{Zoom in of the first and second quadrant arm features in CO from the 207Myr Bc arm model (second row, Fig.\ref{MapLabel}). Observational data is shown in the upper panel for comparison. As the synthetic has stronger emission on average we have increased the emission of the observational data slightly to be on the same scale as the synthetic map in this figure.}
  \label{Q2arms}
\end{figure*}

The second model (Bc) has a slightly lighter disc, and appears a much better fit for many of the \lv{} features. The arms in the second and third quadrant are an especially good reproduction of observations. In Figure \ref{Q2arms} we show a zoom in of this region shown alongside the observational data. The Local, Perseus and Outer arm features are all reproduced and have comparable line-of-sight velocities. The emission is however still somewhat higher than that observed, a problem with all maps produced. The top-down map shows the Local and Perseus arms are a bifurcation of the same arm. The Carina hook structure is reproduced but is present at the incorrect longitude. The tangent point lies upon Vela ($l=100^\circ$) rather than the observed Carina tangent ($l=80^\circ$). This feature can be made to match better by increasing $R_{\rm obs}$, but at the expense of the other arm features. The Inner Ridge of this model is somewhat poorer than the other models. There is a void of emission at approximately $l=20^\circ$ $v_{los}=80$\kms{} where clear emission is seen in observations. The incorrect reproduction of the Inner Ridge is due to the SCC Arm tracing a near-circular arc in the inner disc, which is seen as the steep straight line in the \lv{} map. In the other models the SCC arm clearly ``winds", i.e. has a non-circular shape, so is seen to be angled in \lv{} space.

The second Bc map, created at approximately 100Myrs after the first, appears to be the best reproduction by-eye. The model shows the Carina, Perseus, Outer and Local Arms as well as an Inner Ridge that is aligned similarly to observations. 
This model offers the best reproduction of the Inner Ridge and Carina arm simultaneously. The Carina Arm appears to branch away from the SCC Arm (the source of the Inner Ridge) allowing it to be correctly placed in \lv{} space without causing spurious emission in the inner disc, a problem the symmetric fixed potential models persistently encountered. There also appears to be a 4-armed outer structure, with a strong 2-armed inner structure, which adds weight to the models in the literature that suggest a strong stellar 2-armed component with weaker 4-armed one in the gas/dust and young stars \citep{2000A&A...358L..13D,2009PASP..121..213C}.
Local material is again formed by a spur off the Perseus Arm. The Perseus Arm itself is hard to differentiate from the Local and Outer Arm features, which is the main problem with the model. All arm structures in the second quadrant appear at too shallow velocities, implying $V_{\rm obs}$ is incorrect or that the model rotation curve is too shallow near the Solar Radius. 

Our final model has the lightest disc, and the lowest fit statistic. As with the previous model, there is a good reproduction of the Inner Ridge, Carina Arm and Local material. The Perseus and Outer Arms appear too weak, and there  is a significant amount of emission in the inner disc ($|l|<30^\circ$). While common to all models, this excess emission is especially evident here due to the general flocculent nature of the spiral arms. There is no clear inner disc structure and the many smaller arm features in the inner disc are seen in molecular emission, appearing as a great swath rather than distinct arm features. The flocculent nature is also the cause of the weakness of the Perseus and Outer Arms beyond the Solar radius. It is surprising that the emission features can be well reproduced by a model with seemingly no clear dominant spiral mode (Fourier analysis indicates $m \approx 5$). While the fit statistic indicates this model is a good fit, this is likely because there is little emission seen in the incorrect place, coupled with the correct reproduction of the Inner Ridge, Carina and Local Arms. The arm features however seem too weak in the outer disc for this model to be a correct reproduction of our Galaxy. 

Overall, the fit statistic favours the live bulge-disc model with the lightest disc, Bd. However, the features appear too flocculent in the outer disc, and so we favour the second best fit model, Bc. The Bd model provides an excellent fit to the strongest emission regions, that of the Inner Ridge, but the Bc reproduces other arm features better.

\section{Discussion}
\label{discussion}

\subsection{General results}
In this paper we have shown that a live $N$-body system representing the stellar component of the Galaxy can provide a good match to many of the observed molecular emission features. While not one single model produced all arms perfectly, we believe that with enough initial seeds a match could be found. The Bc model at the later time-stamp in particular compares to the Milky Way remarkably well, with a good reproduction of all arm features and inner emission structure.

The values in Table \ref{SPHNGsimpFits} show that models with the lowest fit statistic ($<0.9$) have a high arm number ($3<N<5$). Models showing 2-armed structures were not readily produced. In the case of the Ba model a 2-armed morphology was highly irregular and appeared to be buckling in the outer disc, beyond the Solar position. 2-armed models have been produced by studies in the literature, but tend to only be so when perturbed by some external body \citep{1972ApJ...178..623T,2000MNRAS.319..377S,2010MNRAS.403..625D,2011MNRAS.414.2498S}. The study of \cite{2004MNRAS.350L..47M} proposed a 2-armed spiral potential that drives a 4-armed structure in the gas. However, in Paper 1 this model was tested and provided a poor match to the \lv{} data.

The spiral arms produced here are transient in nature. Over the order of a Gyr the dominant arm number will be consistent for a given model, but the stellar and gaseous material for each arm reside in that arm until it is sheared apart by differential rotation of the disc. This is at odds with the spiral density wave theory that suggests arm features would prevail over long time periods, with material continually flowing into and out of the potential well of the arms. The fact that the transient arm structures presented here are a better match than the steady waves of the potentials used in Paper 1 (see Sec. \ref{potentialcomp}), and that standing spiral waves are yet to be reproduced in any numerical simulation is evidence that the density wave theory may not be applicable to the Milky Way.

There is however one key component absent from the models shown. The central molecular zone is very broad and featureless, a stark contrast to observations and presumably a result of the lack of an inner bar. 
This is due to the $Q$-barrier caused by the inclusion of the bulge required to match the rotation curve of the Milky Way.
Interestingly all models show a trailing $m=2$ mode in the inner disc. It is possible this is the disc attempting to form a bar but is undermined by the increased stability inherent to the bulge dominated region. While we did manage to produce bars in some of our preliminary simulations, they required a long term evolution of the stellar disc of the order of many Gyrs by which point the strength of the arms is greatly reduced. Additionally the lack of a bulge means the inner rotation curve is a poor match for the Milky Way. The generation of $N$-body bars in Milky Way like models with a co-existing strong spiral structure will be the subject of a future study.

\subsection{Comparison to calculations with fixed stellar potentials}
\label{potentialcomp}
In Paper 1 we performed calculations using a number of fixed analytic potentials to represent the arm and bar components of the Milky Way. A 4-armed model was favoured to match all the arm features, with pitch angles around 15\arcdeg{}, pattern speeds of 20\ps{} and 50-60\ps{} for the arms and bars respectively. While many arm structures could be reproduced, the regular structure made it impossible to create a perfect match. For instance, the Carina arm was impossible to correctly reproduce without placing very strong local emission in the inner Galaxy.

It appears that using a live-stellar distribution provides a much better match for Galactic \lv{} structure than the fixed analytic potentials in many regards, clearly seen by comparing Figure 25 of Paper 1 and Figure \ref{MapLabel} here. The irregular arm structures created by the live stellar system are able to match emission features simultaneously, such as the Carina arm and Inner Ridge, where symmetric logarithmic spiral arms could not. For example, in the models of Paper 1 the second quadrant could be fit by moderate to large pitch angles, whereas a much smaller value was needed to fit the Carina feature. The arm numbers of the best fitting models are similar to those suggested previously, favouring a 4-armed gas structure to best match the observations. The pitch angles are somewhat higher than found in Paper 1 ($18^\circ<\alpha<25^\circ$), and also higher than the standard Milky Way models \citep{2005AJ....130..569V}. High pitch angles are not uncommon in $N$-body simulations \citep{2011IAUS..270..363W,2013A&A...553A..77G} whereas low values are seemingly hard to create without the arms dissipating.
Pattern speeds appear to be a function of radius in nearly all cases, with mean values for arms ranging from $20$\ps{}$<\Omega_p<40$\ps{}, also similar to values found in other studies \citep{2012MNRAS.426..167G,2013ApJ...763...46B} but higher than those found when using analytic potentials. The arms appear material, unlike the steady density waves implied by theories, and material tends to reside in the arms until they shear apart. It may be the case that the Galaxy has no fixed pattern speed, with spiral arms that are also material in nature. Some external galaxies have also been observed to have a pattern speed that decreases with increasing radius \citep{2008ApJ...688..224M,2012ApJ...752...52S}. 

The total strength of CO emission is more in line with that seen in observations (i.e. weaker) than the models of Paper 1, which tended to create extremely bright arm emission structures throughout the arms. The increased amount of inter-arm structures allows for lower strength emission features, whereas analytic potentials tended to create one strong swath of emission tracing the potential minimum. The fit statistic for the emission maps of models is calculated in exactly the same way as in Paper 1 and the best-fitting models here provide a systematically better fit than those with fixed potentials (best fitting values give $\approx 1.05$ for potentials and $\approx 0.95$ for live discs).

There are some drawbacks to this method however. Each model has far too much emission in the inner disc. 
The fixed potential calculations effectively had a hole in the inner disc, which resulted in a large dearth of emission at high velocities inside of $|l|<20^\circ$. High gas density was still seen in the inner disc, but was solely aligned on the $x_2$ orbits of the bar. As such we believe either an inner bar structure is needed to sweep up molecular material in the inner disc, or that gas density is greatly reduced by some other mechanism.

\section{Conclusions}
\label{conclusions}

In this study we have shown simulations of the stellar and gaseous components of the Milky Way. Different spiral morphologies were formed in the stars and gas, with arms appearing transient and material in nature rather than as density waves. The arm number is seen to increase as the disc to halo mass ratio decreases, with arm numbers found to range from $2\leq N \leq5$ similar to those predicted from swing amplification theory. We perform fits to logarithmic spiral features, finding pitch angles in of $18^\circ<\alpha<25^\circ$ and pattern speeds in the range of $20$\ps{}$<\Omega_p<40$\ps{} which decreases with radius rather than maintaining a constant value. Both pattern speed and pitch angle are within the range of values inferred from observations, though in the higher part of this range.

Using the molecular gas in these simulations we then created synthetic \lv{} emission maps. A simple method is used to find a best-fitting time-frame and observer coordinates, which are used to reject some outlying models and provide input parameters for the full radiative transfer maps. We find moderate mass discs (model Bc; $M_d=3.2 \times 10^{10}M_\odot$) with a live bulge-disc component provide a very good match to the observations, with 4-armed spiral patterns that reproduce many of the arm features. These arms provide a better fit than those using fixed potentials and provide a lower fit statistic. The arm features of the Milky Way are thus found to be best-fit by a dynamic and transient disc, displaying a predominantly 4-armed pattern in the gas with a pitch angle of approximately 20\arcdeg{}. 
Some observational studies propose a clearer 2-armed structure (e.g. \citealt{1970IAUS...38..126W,2000A&A...358L..13D,2010ApJ...722.1460S}), whilst some are more favoured towards a 4-armed structure (e.g. \citealt{1976A&A....49...57G,2003A&A...397..133R,2006Sci...312.1773L}), though we note again that our models tend to display 2 arms in the centre of the disc.

The main features of the Milky Way appear to be reproducible with relatively simple physics. By only taking into account the stellar gravitational field we can create a strikingly similar match to the observed features. This implies that, for most of the Milky Way disc at least, the effect of feedback, self-gravity, the bar, and perturber interactions can be considered minimal compared to the stellar gravity.

A clear improvement to the work in this and Paper 1 is the inclusion of additional physics. Aside from hydrodynamics and chemistry, our calculations are relatively simple. The logical next step would be to take the best-fitting models from the analytic potentials and the live stellar systems and include gas self-gravity, stellar feedback or even Galactic scale magnetic fields. It would also be interesting to see whether tidal forces induced by an object of the order of the mass of the LMC could induce a spiral structure resembling what is seen in \lv{} Galactic observations.

\section*{Acknowledgments}

We thank an anonymous referee, whose comments and suggestions improved the paper. We are grateful to Jim Pringle and Daniel Price for helpful comments.
We thank Tom Dame for providing access to the CO longitude-velocity data. 
The calculations for this paper were performed on the DiRAC Complexity machine, jointly funded by STFC and the Large Facilities Capital Fund of BIS, and the University of Exeter Supercomputer, a DiRAC Facility jointly funded by STFC, the Large Facilities Capital Fund of BIS and the University of Exeter. 
ARP was supported by an STFC-funded post-graduate studentship while performing this work and is currently supported by a MEXT grant.
CLD acknowledges funding from the European Research Council for the FP7 ERC starting grant project LOCALSTAR. 
Figures showing SPH density were rendered using \textsc{splash} \citep{2007PASA...24..159P}. 

\bibliographystyle{mn2e}
\bibliography{Pettitt_morphology2.bbl}

\appendix
\section[]{Quantifying arm morphology}
\label{Appx1}

\begin{figure}
\centering
\resizebox{1.0\hsize}{!}{\includegraphics[trim = 10mm 0mm 20mm 5mm]{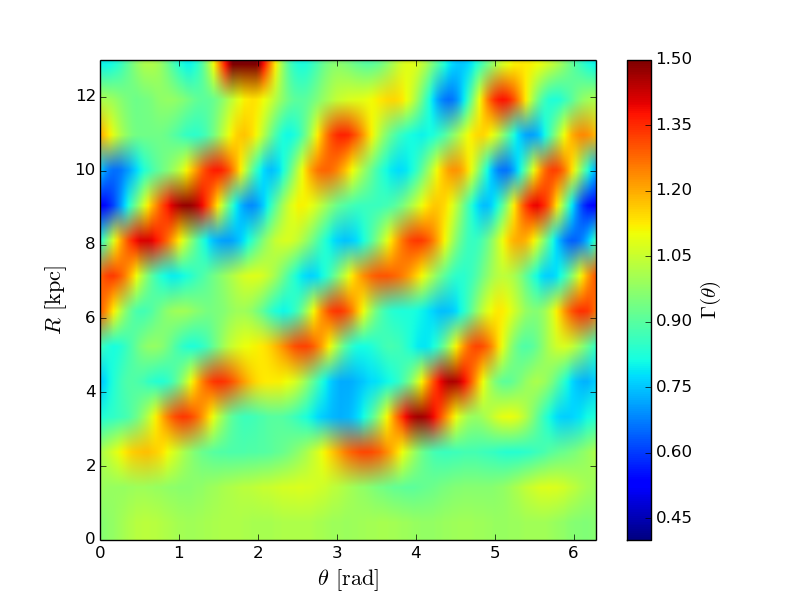}}
  \caption{The stellar component of the Bc model binned into $R-\theta$ space and normalised by azimuthally averaged surface density. The diagonal features show location of spiral arms that are detected by the Fourier decomposition.}
  \label{FourierModes}
\end{figure}

The arm number ($N$), pitch angle ($\alpha$) and pattern speed ($\Omega_p$) are determined using methods similar to those used in pervious studies \citep{2010MNRAS.403..625D,2011ApJ...730..109F,2012MNRAS.426..167G,2013ApJ...766...34D}. An important caveat is that the arm shape is assumed logarithmic and periodic, i.e. with each arm separated by $2\pi/N$. The particles are first binned into a $R$-$\theta$ grid which is then normalised by the surface density of the disc to ensure that inner arm structures do not dominate the fit. This gives the surface density contrast at a given radius: $\Gamma |_R(\theta)=\Sigma(R,\theta)/\Sigma(R)$. An example of which for the Bc model is shown in Figure \ref{FourierModes}.

\begin{figure*}
\includegraphics[trim = 0mm 0mm 0mm 0mm,width=150mm]{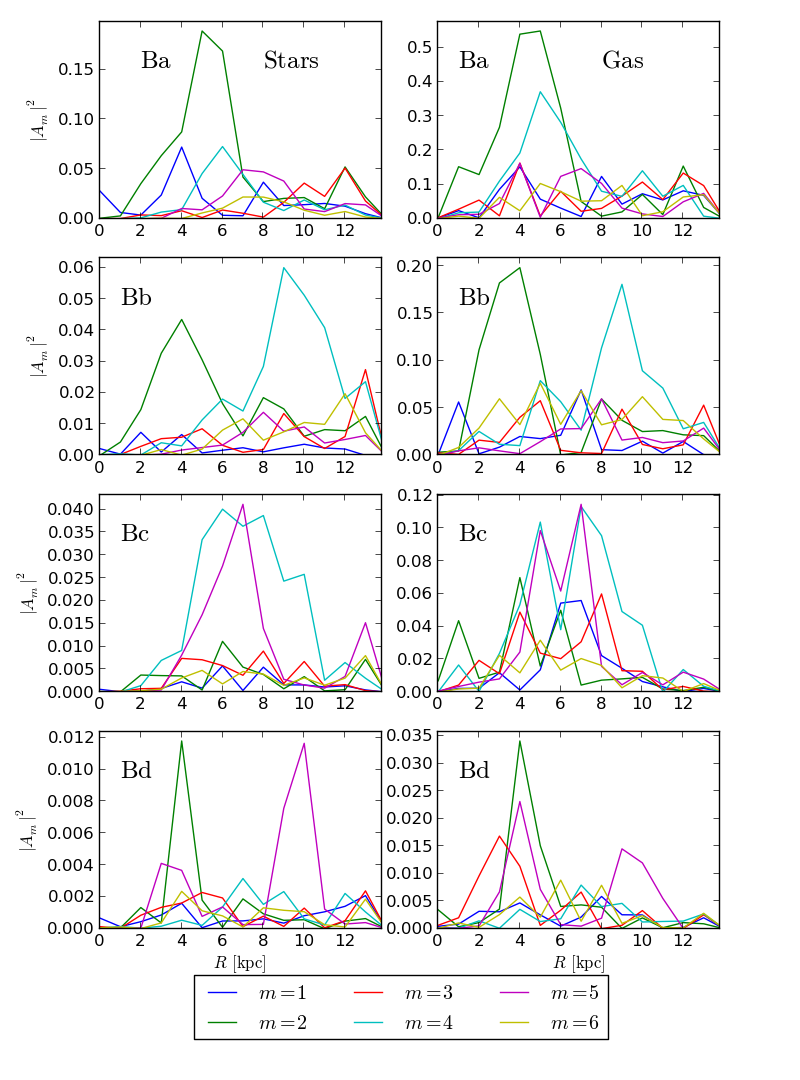}
\caption{The power of individual spiral modes in the range $1\le m\le 6$ for the gaseous and stellar components of models Ba, Bb, Bc and Bd. Similar modes dominate in the gas and stars with higher arm modes dominating the lower mass discs.}
\label{ArmModes}
\end{figure*}

A Fourier transform is then performed on $\Gamma |_R(\theta)$ at different radii, the results of which for the stellar and gaseous components of our four main models (using a live disc and bulge) are shown in Figure \ref{ArmModes}. 
The dominant modes clearly appear to increase with decreasing disc mass, with the $m=2$ mode dominating the Ba model, and the $m=5$ mode gaining significant power in the Bc and Bd models while appearing negligible in Ba and Bb.
The strongest Fourier mode is taken to be the arm number, though in some instances there are two conflicting modes due to the variation as a function of radius. We take the $m=4$ mode in the Bb model, as it is stronger in the stellar component and near the solar position. For Bc the $m=4$ mode is also selected, though the $m=5$ mode displays significant power over a wide radial range in power components. For the Bd model we select the $m=5$ mode. The $m=2$ mode is also strong, but only in the inner disc, whereas the $m=5$ mode is strong in the outer disc yet still peaks in the inner disc.

We then extract the $R$-$\theta$ position of the dominant mode. Depending on the morphology, this may only be done across a certain radial extent, such as the range in which that mode dominates (e.g. from $2{\rm kpc}\leq R \leq6{\rm kpc}$ for the $m=2$ mode in the Bb).
The pitch angle, $\alpha$, of a logarithmic spiral arm is linked to the values of the constant $B$ in the equation
\begin{equation}
\theta = f(R,\alpha)=B\ln{R} +C
\end{equation}
where the pitch angle is $\alpha={\rm arctan}\,B$ and $C$ defines the azimuthal position of the arms. This equation is then fit to the relevant mode using a downhill Nelder-Mead simplex algorithm from the \textsc{scipy} \textsc{python} package and minimising a chi-squared like statistic.

\begin{figure*}
\includegraphics[trim = 0mm 0mm 0mm 0mm,width=150mm]{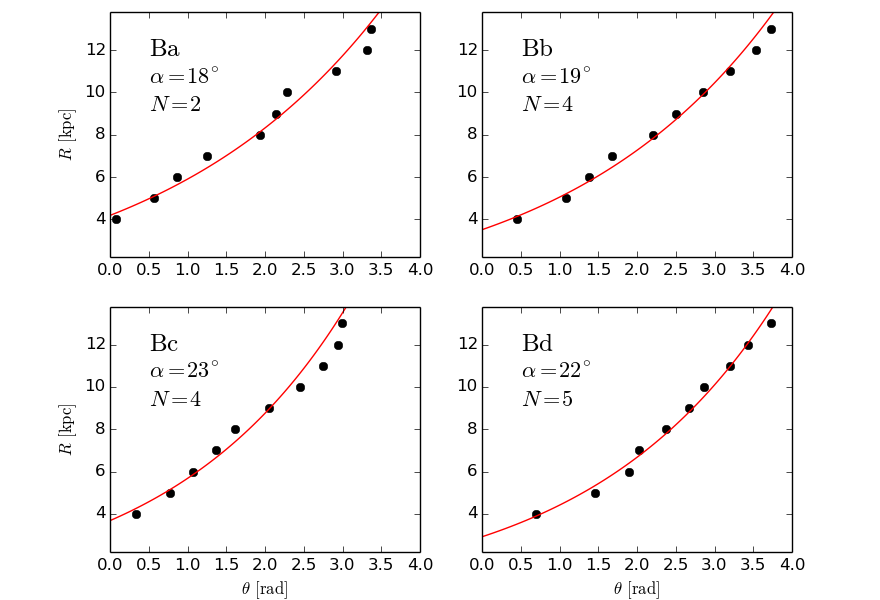}
\caption{Best fitting logarithmic spiral model (red line) to the dominant modes given by \ref{ArmModes} for our live disc-bulge models (black points). The best-fitting pitch angles are given in the top left of each panel.}
\label{ArmFits}
\end{figure*}

Figure \ref{ArmFits} shows the fits to the dominant stellar components of the models in Figure \ref{ArmModes}. Black points trace out the arm and the best fitting logarithmic spiral arm is shown by the red line. The logarithmic spiral appears a very good fit to the arm features in our models, though the weak nature of the Bd model arms leaves a larger margin of uncertainty.

\begin{figure*}
\includegraphics[trim = 0mm 0mm 0mm 8mm,width=140mm]{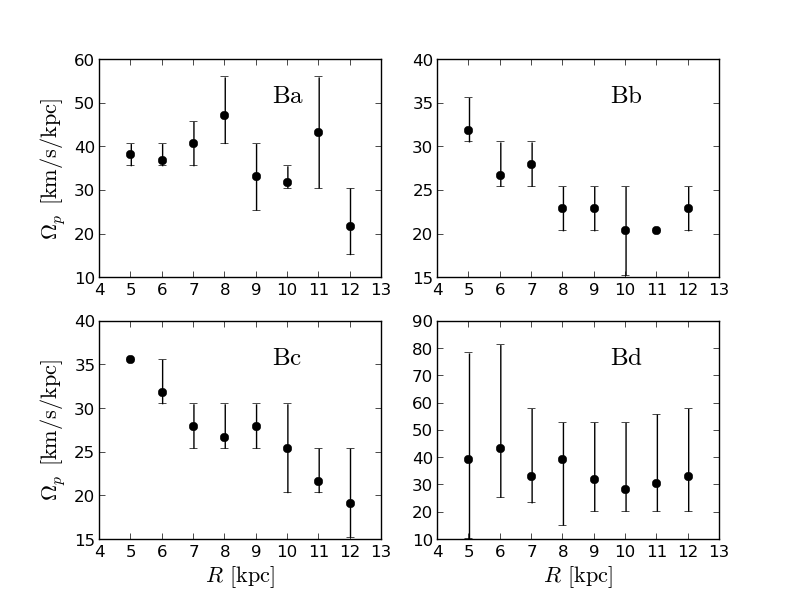}
\caption{Pattern speeds as a function of radius for the arms fit in Figure \ref{ArmFits}. Each point is calculated from a  time interval of 40Myrs, with error bars showing the maxima/minima in pattern speed measured at 10Myr intervals in the 40Myr window.}
\label{ArmSpeeds}
\end{figure*}

We then take subsequent time-steps and perform a similar analysis, this time fitting to only the dominant spiral mode determined from the main time-stamp. Then by simply calculating the offset between arms at different epochs the pattern speed can be calculated as a function of radius. In Figure \ref{ArmSpeeds} we show the pattern speed measured for the models shown in Figures \ref{ArmModes} and \ref{ArmFits}. The individual points are calculated over a range of $\approx$40\,Myrs, with error bars showing the maxima and minima at 10Myr intervals within this window. In all instances a clear decrease of pattern speed with radius can be seen, indicating the arms act as material, rather than wave-like features. 

\bsp
\label{lastpage}
\end{document}